\documentclass[showkeys, amsmath, amssymb,fullpage]{revtex4}

\usepackage{graphicx}
\pdfoutput=1

\begin{document}

\newcommand{\lij}{\lambda_{ij}}
\newcommand{\xij}{x_{i,j}}
\newcommand{\xik}{x_{i,k}}
\newcommand{\xnk}{x_{n,k}}
\newcommand{\pij}{p_{i,j}}
\newcommand{\xnj}{x_{n,j}}
\newcommand{\pnj}{p_{n,j}}
\newcommand{\pnk}{p_{n,k}}
\graphicspath{{/},{figures2/}}
\def\bibsection{\section*{REFERENCES}}

\title{Gravimetric Radar: Gravity-Based Detection of a Point-Mass\\ Moving in a Static Background}

\author{Emmanuel David Tannenbaum}
\email{emanuelt@bgu.ac.il} \affiliation{Department of Chemistry\\
Ben-Gurion University of the Negev\\
Beer Sheva, Israel}

\begin{abstract}

This paper discusses a novel approach for detecting moving massive
objects based on the time variation that these objects produce in
the local gravitational field measured by several detectors.  Such
an approach may provide a viable method for detecting stealth
aircraft, UAVs, cruise, and ballistic missiles.  By inverting a set
of nonlinear algebraic equations, it is possible to use the time
variation in the gravitational fields to compute the mass, position,
and velocity of one or more moving objects.  The approach is
essentially a gravity-based form of triangulation.  Based on
order-of-magnitude calculations, we estimate that under realistic
scenarios, this approach will be feasible if it is possible to
design gravimetric devices that are four to five order of magnitude
more sensitive than current devices.  To achieve such a level of
sensitivity, we suggest designing detectors that exploit a
quantum-mechanical effect known as {\it gravity-induced quantum
interference}.  Furthermore, even if we have a perfect detector, it
will be necessary to determine the magnitude of various atmospheric
disturbances and other sources of noise.

\end{abstract}

\keywords{Gravity, gravity-induced quantum interference, stealth, detector, signal processing}

\maketitle

\section{The Problem of Mass Detection}

Since its introduction during World War II, radar ({\it radio
detection and ranging}) has been the standard method for the
detection of moving objects.  The future efficacy of radar for
military applications is being called into question, as many
countries have become interested in developing weapons systems that
employ various technologies to evade radar detection.  Currently,
only the United States has radar-evading, or ``stealth,'' aircraft
in service, namely the F-22 Raptor and the B-2 Spirit (the first
stealth-capable fighter-bomber, the F-117 Nighthawk, is no longer in
service, while the F-35 Joint Strike Fighter is still under
development).  However, other countries are currently working on
developing stealth-capable aircraft.  Notably, Russia is currently
developing the T-50 fighter jet as a rival to the F-22, with a
planned operational deployment sometime around 2015.

The acquisition and development of stealth technologies by rivals of
the United States presents a potential threat to the United States
and her interests.  Thus, in the near future, it will be necessary
to develop new, stealth-resistant, methods for detecting aircraft
and other moving objects, such as UAVs, cruise and ballistic
missiles. Such detection methods will both provide warning time to
prepare for an impending attack, as well as to identify the location
of a threat, which may then be destroyed with appropriate
counter-measures.

A variety of detection methods other than traditional radar-based
methods are possible. First of all, some stealth technology works by
reflecting the incoming radar beam away from the source. While the
source radar cannot detect the incoming object, using multiple
station radar allows for the reception of this diverted radar
signal, which may then be used to determine the location of the
aircraft via triangulation. The problem with this approach is that
it will not work for stealth aircraft that achieve their
radar-evading capability by absorbing the incoming radar beam, using
so-called {\it radar-absorbing materials}, or RAM. Another approach
relies on detecting the heat signature of an aircraft using infrared
sensors. Once again, the problem with this approach is that modern
stealth aircraft are generally designed to minimize their heat
signatures, make this approach problematic. A third approach
involves using high resolution optical imaging to directly observe
the moving object in the visible spectrum.  While this approach may
be viable for the time being, at least for detection of aircraft
during daytime, there are currently research efforts underway to
design ``cloaking" devices that can bend light around an object and
make the object appear transparent.  Such devices are based on
``meta-materials," whose optical properties can be controlled by
appropriate design of their internal structure. Finally, a fourth
approach involves detecting the sound made by an approaching
aircraft, UAV, or missile. If the object is subsonic, then this
approach is feasible, as the sound wave generated by the object
arrives at the detector before the object itself. However, for
supersonic objects, the sound wave will arrive at the detector after
the object, rendering this approach useless.

Here, we propose an alternative method for the detection of moving
objects.  This method exploits the fact that all massive objects
generate a gravitational field, and that a moving object will lead
to a time-varying gravitational field that can be measured at
various points.  By measuring this time-varying field at a
sufficient number of points, it is possible to obtain the mass,
position, and velocity of the object by solving a system of
nonlinear algebraic equations. This approach has an advantage over
other detection methods, in that, because it is impossible to hide
or shield a gravitational field, this method should be much more
difficult, if not impossible, to counter, than other methods.

There are two main drawbacks and one limitation with this method.
The first drawback is that it requires the ability to detect
gravitational fields that are four to five orders of magnitude
weaker than what is possible with current gravimetric devices. The
second drawback is that, even with a perfect detector, the
gravitational signal generated by the moving object of interest may
be masked by effects such as atmospheric disturbances and clutter
due to the random motion of various other objects (e.g. cars,
animals, etc.).

A potential limitation of this method is that it may not work well
for ships and submarines that displace a mass of water equal to
their own mass. The reason for this is that a variation in the
gravitational field is generated by variations in the mass density
distribution. If a moving ship or submarine simply displaces an
equal mass of water, then the mass distribution may not change
sufficiently to lead to a detectable signal.  Thus, this mass
detection method, if feasible, is likely only to be applicable to
massive objects that travel on the ground or in the air.

This paper contains the basic theory underlying a gravity-based
approach for mass detection.  We will discuss various ways to deal
with anticipated drawbacks of this method.  In particular, we will
propose an initial set of studies to determine whether the method is
at all feasible. Therefore, the present work has the nature of an
entire research program. A full journal version with details of the
relevant physical background may be found in \cite{etannenbaum}.

\section{Gravity-Based Detection of Moving Objects: Theory}

According to Newton's Universal Law of Gravitation, two
point-objects of mass $ m_1 $ and $ m_2 $ interact through a
gravitational force of magnitude $ F = G m_1 m_2/r^2 $, where $ r $
is the distance between the objects, and $ G $ is the {\it
gravitational constant}, which in SI units is $ 6.67300 \times
10^{-11} \rm[meters^3 kg^{-1} s^{-2}] $.  The force is purely
attractive, so that it is directed along the line connecting the two
objects.

For more than two point masses, the gravitational force acting on a
given mass is simply the sum of all the forces due to the
interactions with each of the other masses.  Generalizing to a mass
distribution, we obtain that the gravitational field at a given
point is given by the integral,
\begin{equation}
\vec{g}(\vec{x}, t) = G \int_{R^3} d \vec{x}' \frac{\rho(\vec{x}', t) (\vec{x}' - \vec{x})}{||\vec{x}' - \vec{x}||^3}
\end{equation}
where $ \vec{g}(\vec{x}, t) $ is the gravitational field at the point $ \vec{x} $ at the time $ t $, and $ \rho $ denotes the mass distribution function.  Note that we include an explicit time dependence into our formula, since our mass distribution may be time-dependent, which will then generate a time-varying gravitational field.

Now, we may partition the mass distribution into a ``background,'' consisting of the Earth and atmosphere, and the ``object,''
consisting of the object, or objects, to be detected.  We assume that the time-variation of the mass distribution is due to the
object component, and so we may write,
\begin{equation}
\vec{g}(\vec{x}, t) = \vec{g}_{background}(\vec{x}) +
\vec{g}_{object}(\vec{x}, t)
\end{equation}
Differentiating both sides of the equation with respect to time, we obtain that,
\begin{equation}
\frac{\partial \vec{g}}{\partial t} = \frac{\partial
\vec{g}_{object}}{\partial t}
\end{equation}
This equation implies that the time-variation of the local gravitational field is due to the object alone.  For a single moving
point-object of mass $ M $, located at $ (x, y, z) $, and, assuming a detector located at the coordinates $ (x_i, y_i, z_i) $, we have,
\begin{widetext}
\begin{eqnarray}
&   & g_{object, x}(x_i, y_i, z_i, t) = G M \frac{x - x_i}{[(x -
x_i)^2 + (y - y_i)^2 + (z - z_i)^2]^{3/2}}
\nonumber \\
&   & g_{object, y}(x_i, y_i, z_i, t) = G M \frac{y - y_i}{[(x -
x_i)^2 + (y - y_i)^2 + (z - z_i)^2]^{3/2}}
\nonumber \\
&   & g_{object, z}(x_i, y_i, z_i, t) = G M \frac{z - z_i}{[(x -
x_i)^2 + (y - y_i)^2 + (z - z_i)^2]^{3/2}}
\end{eqnarray}
\end{widetext}

Differentiating with respect to time, we obtain,
\begin{widetext}
{\footnotesize\begin{eqnarray} \dot{g}_{x}(x_i, y_i, z_i, t) = G
\frac{-2 (x - x_i)^2 p_{x} + (y - y_i)^2 p_{x} + (z - z_i)^2 p_{x} -
3 (x - x_i) (y - y_i) p_{y} - 3 (x - x_i) (z - z_i) p_{z}}{[(x -
x_i)^2 + (y - y_i)^2 + (z - z_i)^2]^{5/2}} \label{eq:5}
\\
\dot{g}_{y}(x_i, y_i, z_i, t) = G \frac{-2 (y - y_i)^2 p_{y} + (z -
z_i)^2 p_{y} + (x - x_i)^2 p_{y} - 3 (y - y_i) (z - z_i) p_{z} - 3
(y - y_i) (x - x_i) p_{x}}{[(x - x_i)^2 + (y - y_i)^2 + (z -
z_i)^2]^{5/2}} \nonumber
\\
\dot{g}_{z}(x_i, y_i, z_i, t) = G \frac{-2 (z - z_i)^2 p_{z} + (x -
x_i)^2 p_{z} + (y - y_i)^2 p_{z} - 3 (z - z_i) (x - x_i) p_{x} - 3
(z - z_i) (y - y_i) p_{y}}{[(x - x_i)^2 + (y - y_i)^2 + (z -
z_i)^2]^{5/2}} \nonumber
\end{eqnarray}}
\end{widetext}
where $ \vec{p} =(p_x, p_y, p_z) := (M \dot{x}, M \dot{y}, M
\dot{z}) $ is the momentum of the object.

Note that, if we know the position and momentum coordinates of the
object, then we can obtain the velocity by differentiating the
position, and from there we can compute the mass.  Therefore, the
motion of the object is completely characterized by six coordinates.
Since each detector provides three pieces of information about the
object, namely, the time variation of the gravitational field along
each of the coordinate axes, with at least two detectors it is
possible to solve a system of six nonlinear algebraic equations and
determine the position and momentum of the object. As we will
discuss later in this paper, for certain technical reasons we may
choose to only make use of the $ x $ and $ y $ components of the
gravitational field, in which case at least three detectors will be
necessary.

The above approach is in principle readily generalizable to the
detection of an arbitrary number of $ N $ point-objects. Since each
object is characterized by six parameters, $ N $ objects are
characterized by $ 6 N $ parameters. If we use all the components of
the gravitational field at each detector, then since each detector
provides us with three measured parameters, we require at least $ 2
N $ detectors.  If instead we only use the $ x $ and $ y $
components of the gravitational field at each detector, then we
require at least $ 3 N $ detectors. In this paper, since we are
considering 1 point-object, and we will use all the components of
the gravitational field at each detector, we will analyze the case
of 2 detectors.

\section{Numerical Solution of the Single Object Case}

We will provide exact details of the solution of
Equations~(\ref{eq:5}) for the case of a single object and a single
object in Sections~\ref{sec:appendixB} and \ref{sec:appendixC}. Here
we give an overview of these methods for the convenience of the
reader.

We may solve the nonlinear system of equations~\ref{eq:5} using
Newton-Raphson Iteration, or Newton's Method. Because the object is
moving along a continuous path, denoted $\vec{x}(t),$ then, if we
know the position and momentum of the object at some time $t$, we
may use this information as the initial guess for determining the
position and momentum of the object at time $t + \Delta t$. If
$\Delta t$ is sufficiently small, then the object's position and
momentum will have changed by a sufficiently small amount that this
initial guess will converge to the object's position and momentum at
time $t + \Delta t.$

Continuing in this way, the object's position and momentum at one
time may be used as the initial guess to obtain the object's
position and momentum at some time in the near future. The result is
that Newton's Method may be used to readily track the movement of
the object. However, this requires some initial position and
momentum for the object, i.e., it is necessary to first acquire the
object. With object, or target, acquisition, we do not have any kind
of \emph{a priori} information on the object that can be used as a
good initial guess for Newton's Method. Thus, target acquisition is
much more difficult than target tracking, because the initial guess
is far likelier to significantly deviate from the true position and
momentum of the object. To get around this problem, and to enable
target acquisition with a fairly ''bad''  initial guess, we use a
method that we call \emph{Newton-Raphson Iteration with Solution
Deformation}, which is discussed in Section~\ref{sec:appendixC}.

\begin{figure}
\centering
\includegraphics[width=4.10in]{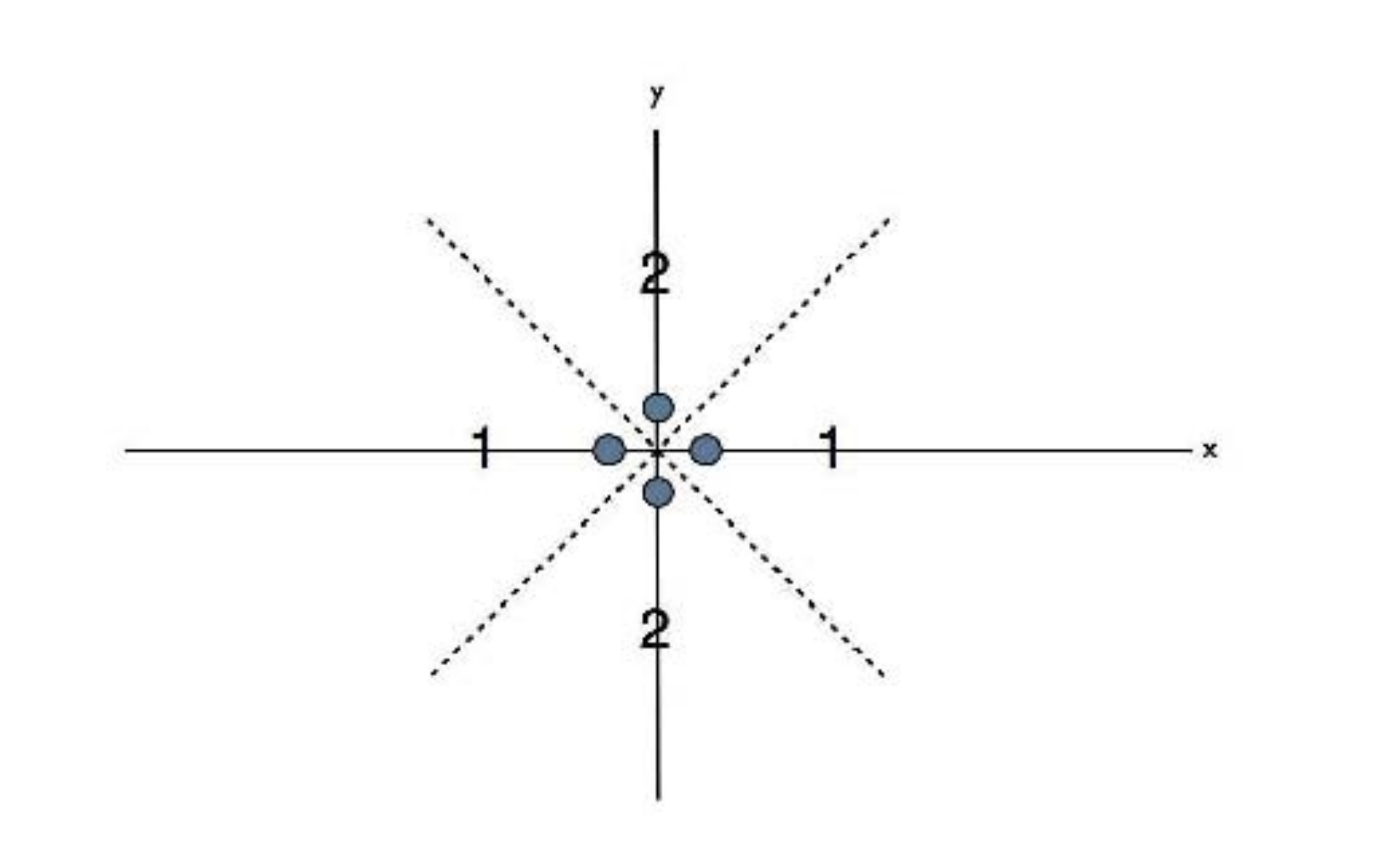}
\caption{Regions of convergence for detectors placed at $(\pm
d/2,0,0)$ and $(0;\pm d/2,0)$. For the first set of detectors, the
region of convergence is given by $\{(x,y,z): \, |y| \le |x|\}$. For
the second set of detectors, the region of convergence is given by
$\{(x,y,z): \, |y|>|x|\}$. \label{fig:figure1} }
\end{figure}

We place detectors at $(\pm d/2,0,0)$ and $(0, \pm d/2,0),$ where
$d$ denotes the detector spacing. The reason for this is that we
have found that target acquisition using our modified Newton's
Method only occurs if the target is located within a region defined
by the line connecting a given detector pair. For the pair of
detectors located at $(\pm d/2,0,0),$ this region is defined by
$\{(x,y,z): \, |y| \le |x|\}$ which is illustrated in
Figure~\ref{fig:figure1}. Therefore, in order to obtain target
acquisition independently of the initial target location, we also
place detectors at $(0;\pm d/2,0),$ which has a region of
convergence given by $\{(x,y,z): \, |y|>|x|\}$, which is also
indicated in Figure~\ref{fig:figure1}.

For each pair of detectors, we consider two initial guesses,
corresponding to the two sectors defining the region of convergence
for the detector pair. For the pair of detectors at $(\pm d/2,0,0)$,
we consider the initial guesses $(x_1, x_2, x_3, p_1, p_2, p_3) =
(x_0,0,0,-p_0,0,0),\, (-x_0,0, 0,p_0, 0,0),$ where the first initial
guess is used for objects in the set  $\{(x,y,z): \, x>0, |y| \le
x\}$ while the second initial guess is used for objects in the set
$\{(x,y,z): \, x<0, |y| \le -x\}.$

Similarly, for the pair of detectors at $(0,\pm d/2,0)$, we consider
the initial guesses $(0, x_0,0,0,-p_0, 0)$ and $(0,-x_0, 0,0,p_0,
0).$ Note that this leads to a set of four initial guesses with
which to acquire the target, at two guesses per pair of detectors.
The algorithm cycles through the four initial guesses and associated
detector pairs until the target is acquired.

If the acquisition fails after going through the entire object
track, the algorithm stops and indicates that this is the case.
Otherwise, the procedure begins to track the object using the
detector pair with which the object was acquired, and indicates the
time at which the target was acquired. Target tracking continues
until the program cycles through the entire input object track, or
until target acquisition is lost. At this point, our proposed scheme
attempts to re-acquire the target using the target acquisition
component of our algorithm. If target re-acquisition is achieved,
the algorithm indicates when this happened, and continues with
target tracking. Target re-acquisition and tracking continues as
necessary until the procedure has run through the entire object
track.

The object track is input as a series of waypoints, specifying the
location of the object at various time intervals. In the computer
implementation of our proposed methodology, the user specifies a
certain number of waypoints, and provides the $(x, y, z)$
coordinates for each waypoint. The user also specifies a time
interval $\Delta t$ and a time interval number $\rm{Num}_{\Delta
t}$, so that the total time between waypoints is $T_{\rm{WayPoint}}
= \rm{Num}_{\Delta t} \times \Delta t.$ We assume a constant
velocity between waypoints, so that the successive locations of the
waypoints, as well as the value of $T_{\rm{WayPoint}}$ may be
readily used to compute the velocity of the object. Combined with
the user-input object mass, this allows us to compute the momentum
of the object between waypoints.

\section{Simulations} \label{sec:simulations}

\begin{figure}
\centering
\includegraphics[width=4.10in]{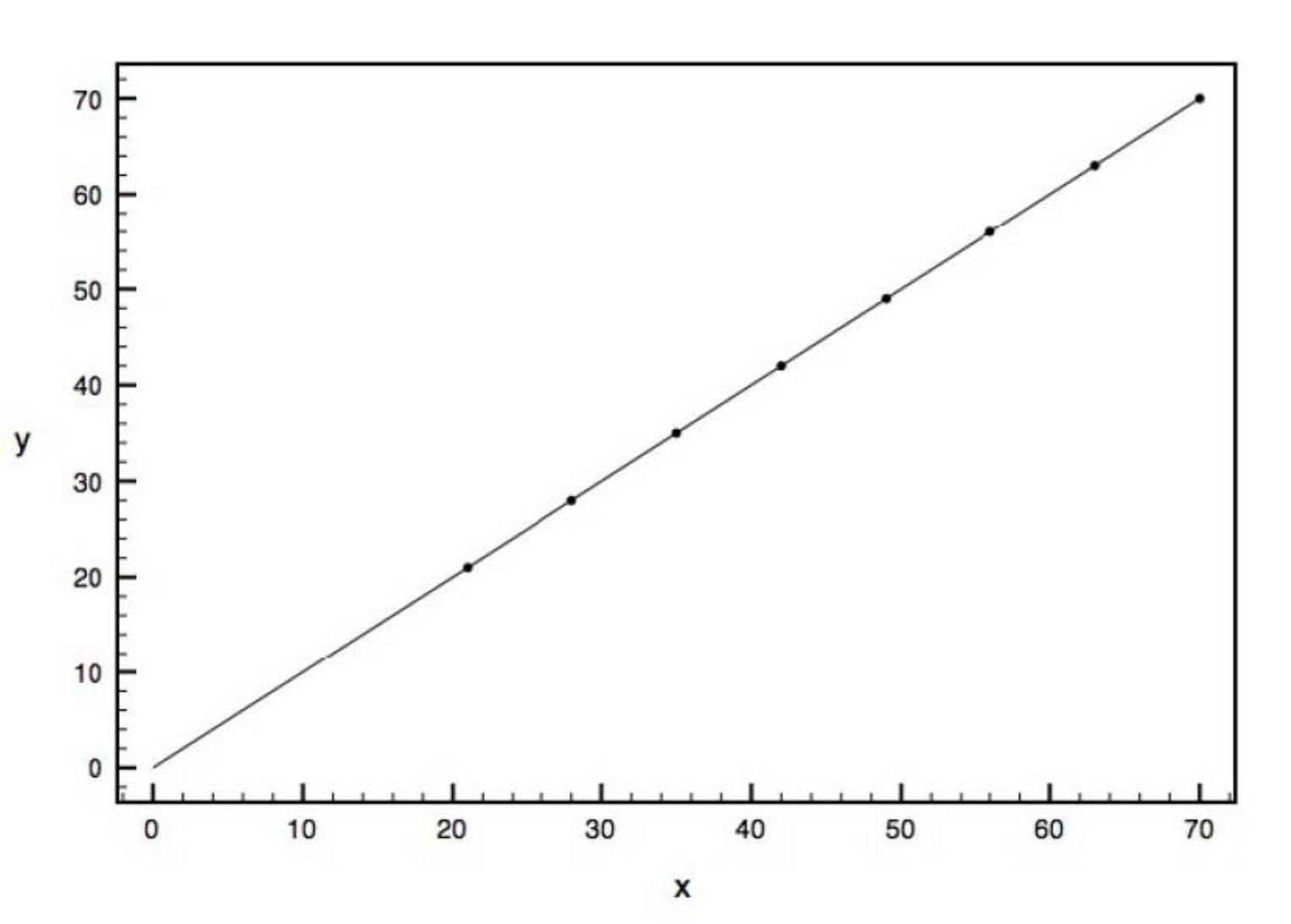}
\caption{Comparison of actual object tracks (solid line) versus
computed object track (dots) for the ``raid'' scenario. Object mass
= 100, with two waypoints, at $(70,70,10)$ and $(0, 0, 10).$ We took
a value of $\Delta t = 0.01$ and $T_{\rm{WayPoint}} = 10,$ giving an
object velocity of 990, and a momentum of $99,000.$ We took a
detector spacing $d = 1,$ and initial guess parameters $x_0 = 50,$
$p_0 = 10,000.$ We used 100 sub-intervals for the solution
deformation implementation of the target acquisition algorithm. The
target was immediately acquired at its initial position, but target
acquisition was lost at $t = 0.08.$ \label{fig:figure2}}
\end{figure}

\begin{figure}
\centering
\includegraphics[width=4.10in]{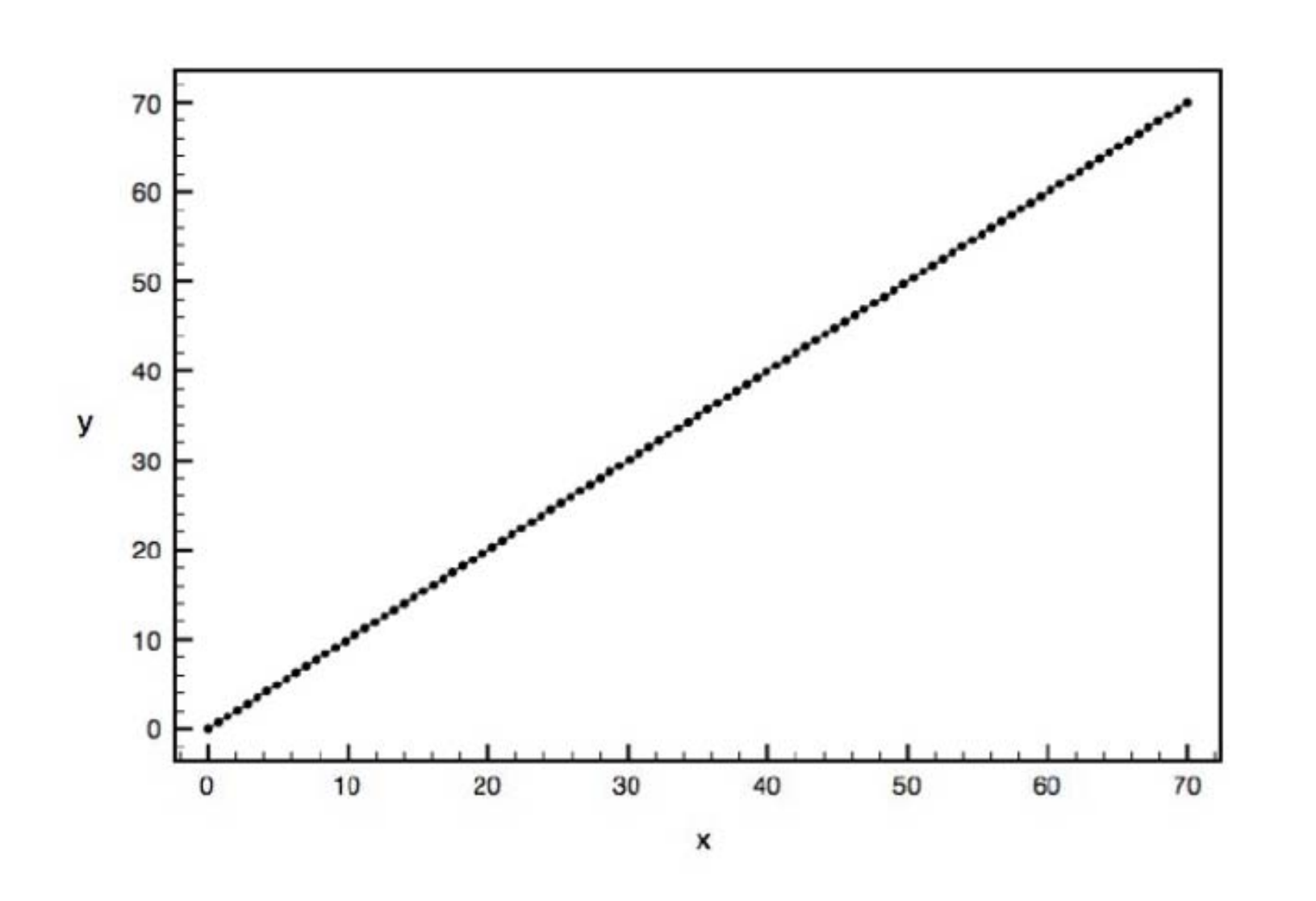}
\caption{Comparison of actual object track (solid line) versus
computed object track (dots) for the ''raid'' scenario. All
parameters are identical as for Figure~\ref{fig:figure2}, except
that now we took $\Delta t = 0.001$ and $Num_{\Delta t} = 100.$ Here
target acquisition was immediate and never lost.
\label{fig:figure3}}
\end{figure}

\begin{figure}
\centering
\includegraphics[width=4.10in]{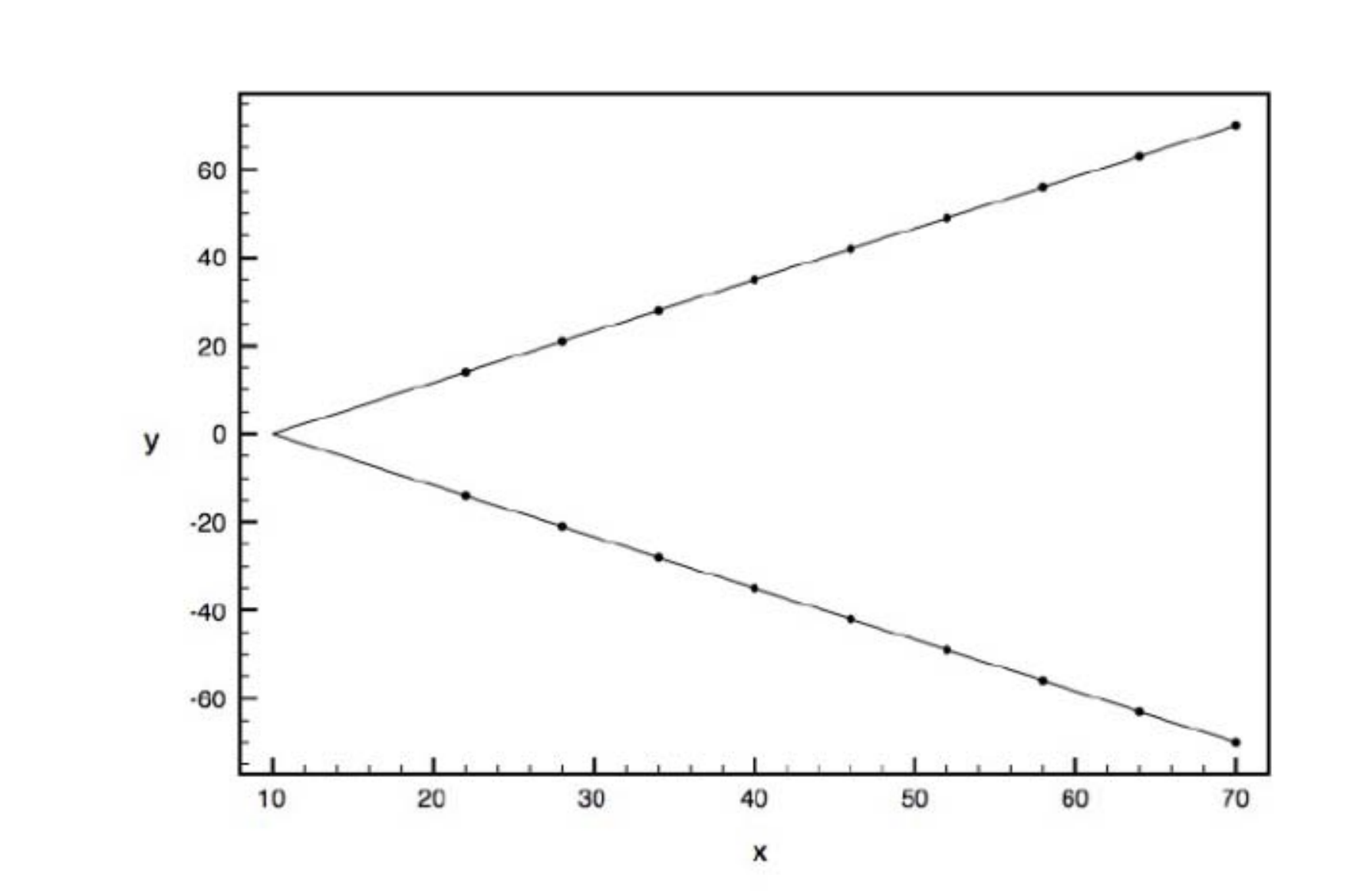}
\caption{Comparison of actual object track (solid line) versus
computed object track (dots) for the ``reconnaissance'' profile.
Object mass = 100, with three waypoints, at $(70,70,10)$, $(10, 0,
10),$ and $(70,-70,10).$ We took a value of $\Delta t = 0.01$ and
$\rm{Num}_{\Delta t} = 10$. We took a detector spacing of $d = 1,$
and initial guess parameters $x_0 = 50,$ $p_0 = 10, 000.$ We used
100 sub-intervals for the solution deformation implementation of the
target acquisition algorithm. The target was immediately acquired at
its initial position. Target acquisition was lost at $t = 0.09$, but
re-acquired at $t = 0.12.$\label{fig:figure4}}
\end{figure}

\begin{figure}
\centering
\includegraphics[width=4.10in]{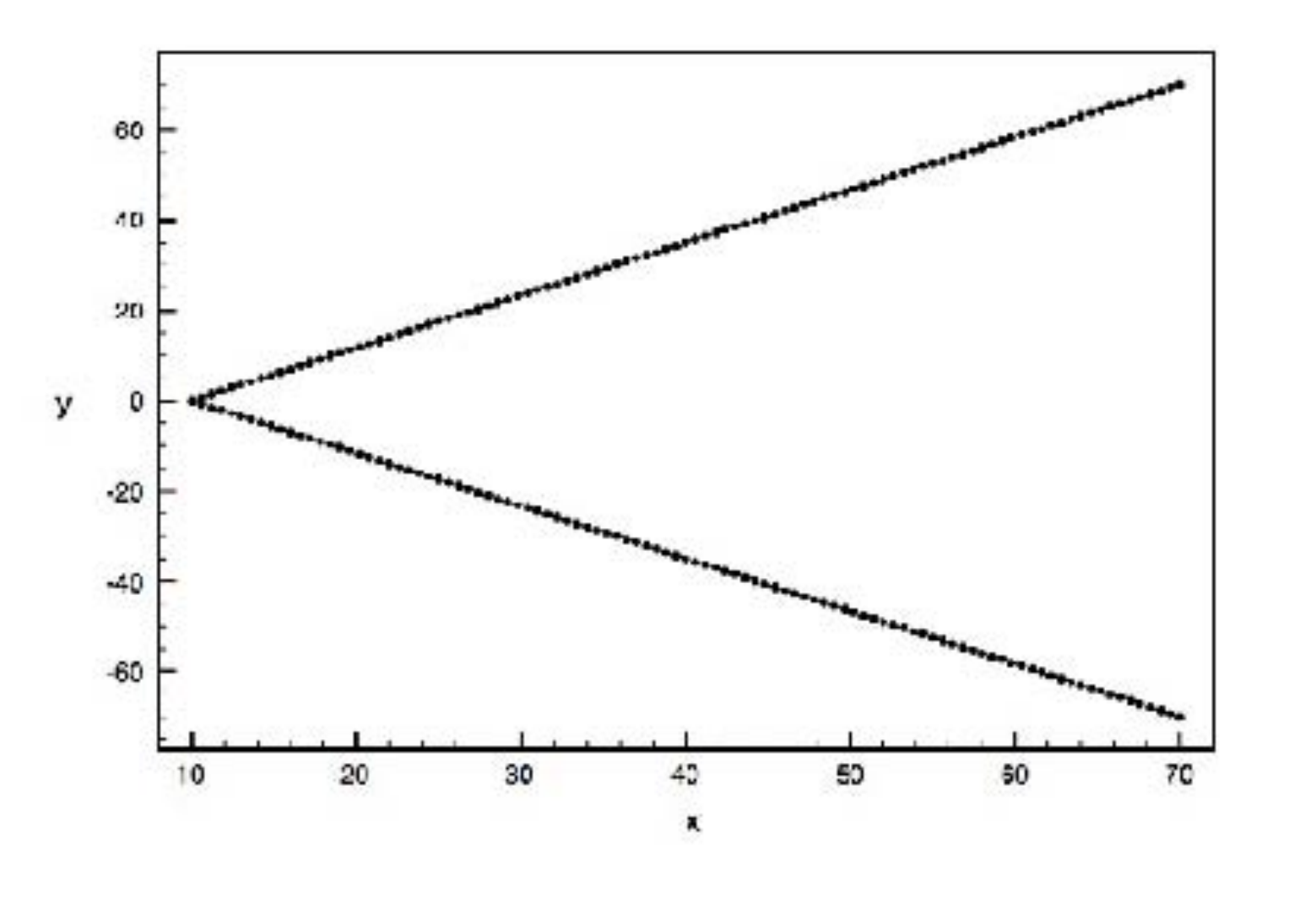}
\caption{Comparison of actual object track (solid line) versus
computed object track (dots) for the ``reconnaissance'' profile. All
parameters are identical as for Figure~\ref{fig:figure4}, except
that now we took $\Delta t = 0.001$ and $\rm{Num}_{\Delta t} = 100.$
Here target acquisition was immediate and never lost.
\label{fig:figure5}}
\end{figure}

We consider three scenarios: In the first case, an object starts at
some distance $R,$ with some altitude $h,$ and heads directly for
the origin at a constant velocity $v.$ The initial coordinates of
the object are given by $(R\cos \theta, R\sin \theta, h),$ and the
velocity vector of the object is $(-v \cos \theta,-v \sin \theta,
0).$ This flight path corresponds to a ``raid'' scenario on a target
at the origin. The results of our target acquisition and tracking
algorithm are shown in Figures~\ref{fig:figure2} and
\ref{fig:figure3} (parameters are provided in the figure captions).
In the second case, an object first moves in a straight line toward
a point near the origin, and then moves away, in a ''zigzag'' flight
pattern, corresponding to a ''reconnaissance'' profile. The results
of our target acquisition and tracking algorithm is shown in
Figures~\ref{fig:figure4} and \ref{fig:figure5} (parameters are
provided in the figure captions).

\begin{figure}
\centering
\includegraphics[width=4.10in]{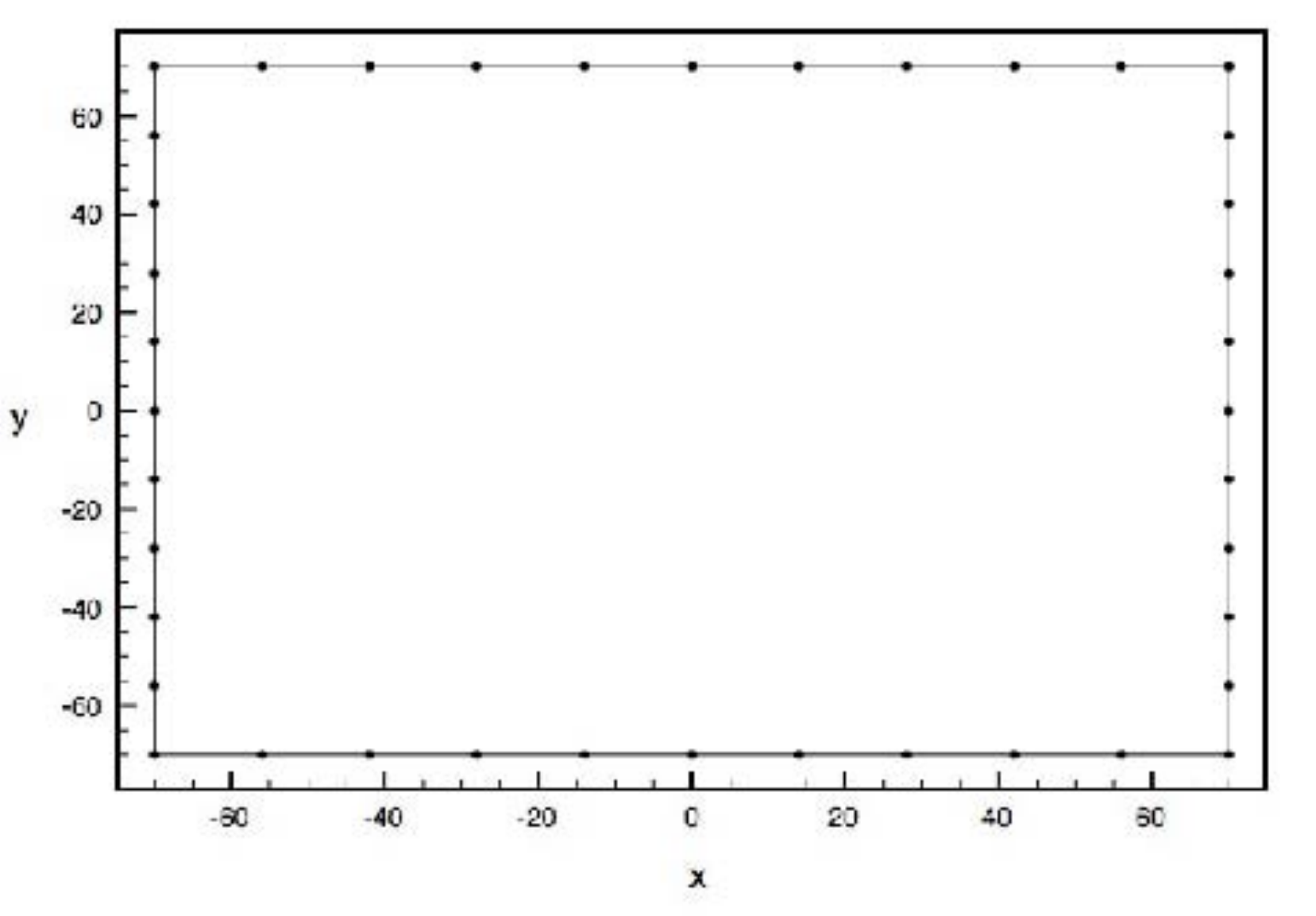}
\caption{Comparison of actual object track (solid line) versus
computed object track (dots) for a square flight profile. Object
mass = 100, with five waypoints, at $(70,70,10),$ $(-70,70,10),
(-70,-70,10), (70,-70,10),$ and $(70,70,10).$ We took a value of
$\Delta t = 0.01$ and $\rm{Num}_{\Delta t} = 10.$ We took a detector
spacing of $d = 1,$ and initial guess parameters $x_0 = 50,$ $p_0 =
10,000.$ We used 100 sub-intervals for the solution deformation
implementation of the target acquisition algorithm. The target was
immediately acquired at its initial position, and tracked over the
entire time of the object track.\label{fig:figure6}} \end{figure}

\begin{figure}
\centering
\includegraphics[width=4.10in]{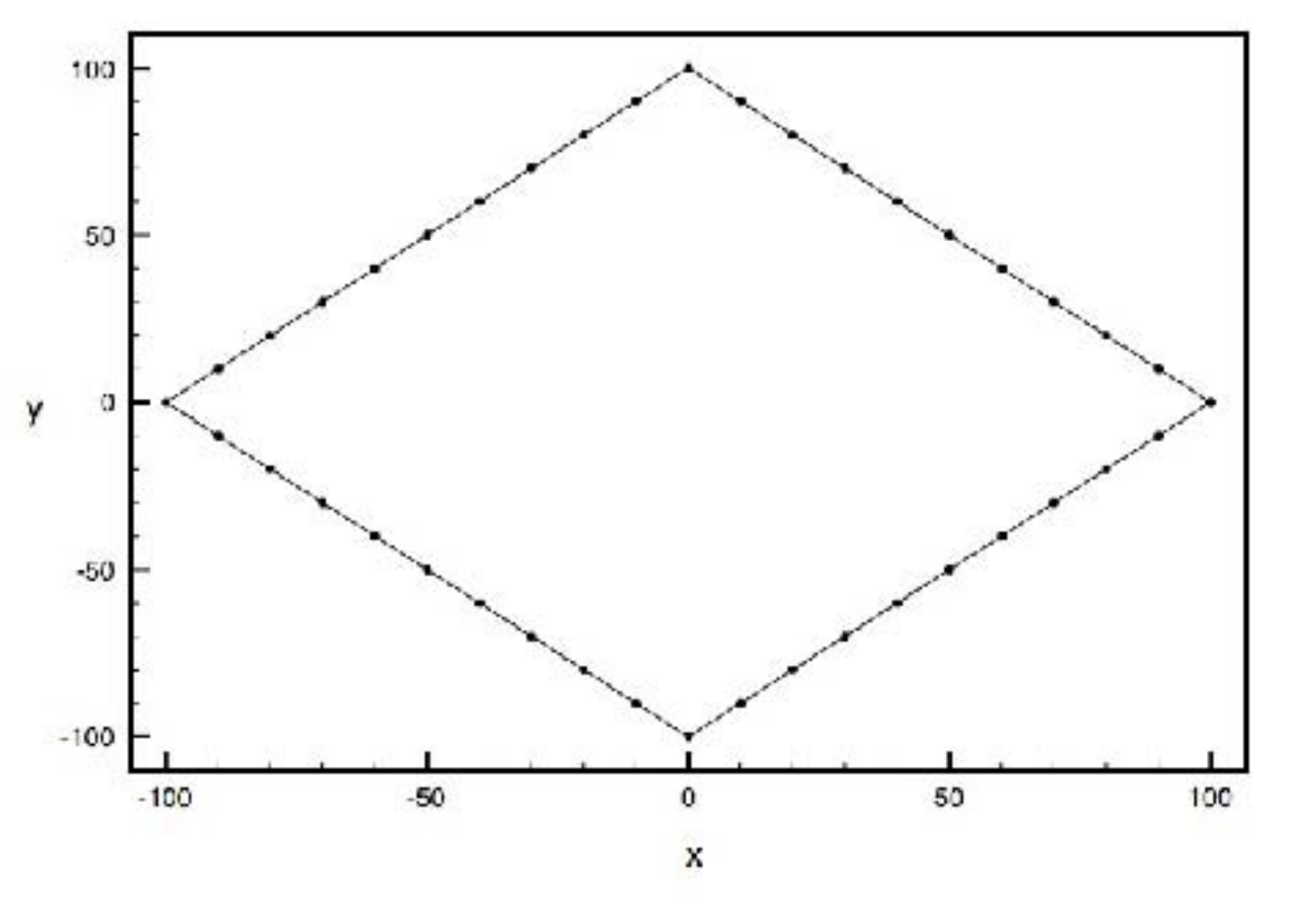}
\caption{Comparison of actual object track (solid line) versus
computed object track (dots) for a diamond flight profile. Object
mass = 100, with five waypoints, at $(100, 0,10),$ $(0,100,10),$
$(-100, 0,10),$ $(0,-100,10),$ and $(100,0,10).$ We took a value of
$\Delta t = 0.01$ and $\rm{Num}_{\Delta t} = 10.$ We took a detector
spacing of $d = 1,$ and initial guess parameters $x_0 = 50,$ $p_0 =
10,000.$ We used 100 sub-intervals for the solution deformation
implementation of the target acquisition algorithm. The target was
immediately acquired at its initial position, and tracked over the
entire time of the object track.\label{fig:figure7}}

\end{figure}

Finally, in the third case, we consider an object that moves in a
square or diamond flight pattern around the origin with constant
speed. The results of our target acquisition and tracking algorithm
is shown in Figures~\ref{fig:figure6} and \ref{fig:figure7}
(parameters are provided in the captions).

\section{Gravity-Induced Quantum Interference} \label{sec:gravity}

In order for gravity-based detection to emerge as a practical method
for detecting moving objects, it will be necessary to develop
devices that can detect gravitational fields several orders of
magnitude weaker than what is possible with current instruments. An
accessible discussion about about quantum interference may be found
in \cite{Andrews}. To illustrate, suppose we wish to detect an
object with a mass of 100[metric tonnes] = $10^5$[kg], at a distance
of 100[km] flying at a speed of $10^3$[km/hr] $\equiv$ 278[meters/s]
(these parameters are based on those from an aircraft such as the
B-2 Spirit). Assuming that this object is headed directly toward the
detector, such an object will lead to a local fluctuation in the
gravitational field of approximately $7 \times 10^{-16}$
[meters/$\rm{s}^2$], and a time variation in the local gravitational
field of approximately $4 \times 10^{-18}$[meters/$\rm{s}^3$]. The
most sensitive gravimetric device to date is the superconducting
gravimeter, which is capable of measuring changes of
$10^{-11}$[$\rm{meters}/s^2$] in the local gravitational field
\cite{shlomi}. This is about four orders of magnitude larger than
what is required to detect a moving object with the characteristics
given above. Clearly, then, to make our gravity-based detection
method practical, we will need to develop gravimetric devices that
are far more sensitive than what is currently available. One
possible approach for the development of a gravimetric device with
the required sensitivity relies on a quantum-mechanical effect known
as \emph{\textbf{gravity-induced quantum interference.}}

\begin{figure}
\centering
\includegraphics[width=4.10in]{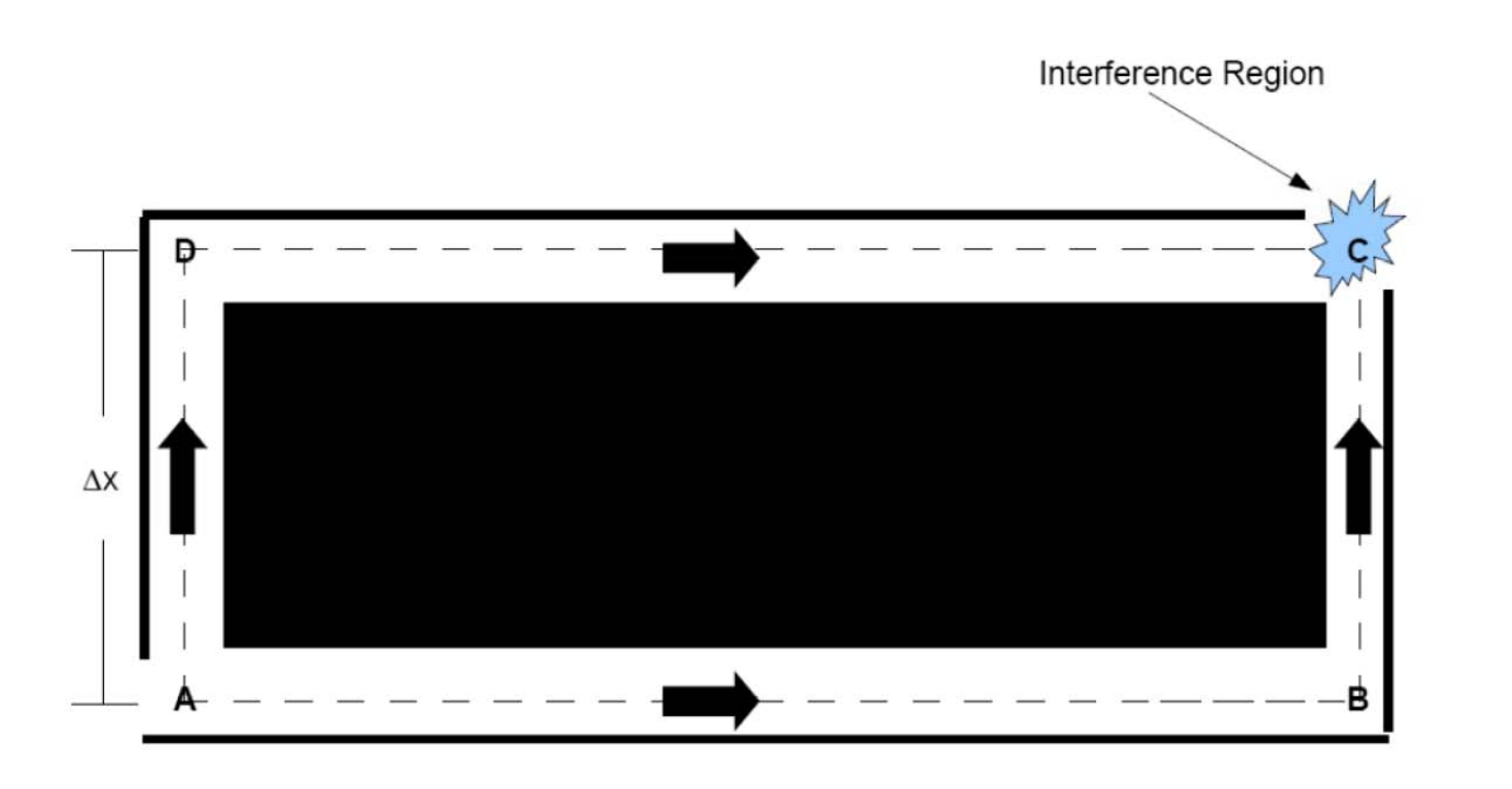}
\caption{A particle moving along path $ADC$ experiences a phase
shift relative to the particle moving path $ABC$, due to a
difference in potential energy between paths $DC$ and $AB$
, given by
$ma \Delta x$, where $a$ denotes the local acceleration due to the
local gravitational field. \label{fig:figure8} }
\end{figure}

Gravity-induced quantum interference is an interference phenomenon
that occurs when a particle interferes with itself after traveling
along two paths with differing potential energies in a gravitational
field. If a particle is introduced into a waveguide as illustrated
in Figure~\ref{fig:figure8}, then the particle may either travel
first along the bottom path (path $AB$) and then up toward the
interference region (path $BC$), or first along path AD and then
toward the interference region (path $DC$).

Because of the potential energy difference between paths $AB$ and
$DC,$ the classical momentum of a particle moving along the two
paths differs, resulting in an accumulated phase difference. The
result is that, when the particle interferes with itself in the
interference region at point $C,$ the probability density is
characterized by an interference pattern of length $L = {hv}/(ma
\Delta x),$ with a time-variation given by $\dot{L} = -L^2 m \Delta
x \dot{a}/(hv).$ Here, $h = 6.626068 \times
10^{-34}$[$\rm{meters}^2$kg/s] is Planck's constant, $v$ is the
classical velocity of the particle moving through the detector, $m$
is the particle mass, $a$ is the gravitational acceleration along
the width of the detector (paths $AD, BC$), and $\Delta x$ is the
height of the detector. Now, taking $\Delta x = 1$[meter], and using
parameters obtained from our sample point object modeled after the
B-2 Spirit, we obtain that $L = 10^{18}(v/m)$[meters] (taking into
account the units of the other factors as well), and $\dot{L} = 5.5
\times 10^{15}(m/v)$[meters/s]. If $m/v = 10^{18}$[kgs/meters], we
obtain that $L \equiv 1[meter],$ and $\dot{L} \equiv 5.5$[mm/s].
These parameters are within readily measurable limits, so that, if
we can design a gravity-based interferometer that produces an
interference effect characterized by these parameters, then it
should be possible to design a gravimetric device with the necessary
sensitivity.

One approach for obtaining the desired mass to velocity ratio is to
take an object with a mass of $10^6$[amu] moving at $1.7$[mm/sec].
Such an object has a mass comparable to that of a virus. While it
may seem counter-intuitive or difficult to exploit quantum
interference effects with such massive objects, it should be noted
that the double-slit diffraction experiment has been pushed
considerably beyond electrons to small molecules. There is currently
active research on creating quantum superpositions of even more
massive objects such as viruses, and perhaps even as large as 1
millimeter. Of course, it is necessary to cool such objects to near
absolute zero in order to ensure that they are in their ground
quantum state, thereby preventing a phenomenon known as decoherence
from destroying quantum superposition effects. Furthermore, in order
to exploit these effects to create a usable device, it will be
necessary to create a steady particle beam out of these
comparatively massive objects. Clearly then, while the design of
ultra-sensitive gravimeters based on gravity-induced quantum
interference is not necessarily an idea based in science fiction, it
is nevertheless at the outer edge of our current technological
capabilities. A second possibility would be to use what is known as
a coherent matter wave, or ``matter laser,'' generated from a
Bose-Einstein condensate. A Bose-Einstein condensate, or BEC, is a
phase of matter whereby all of the particles are in the ground
energy state. Because the particles of a BEC are in the same quantum
state, the BEC exhibits strong quantum superposition effects at
macroscopic scales.

A BEC fluid cannot be treated using classical fluid mechanics,
rather, a purely quantum-mechanical approach must be adopted. In
analogy with coherent light that is used to create a laser,
researchers are interested in using BECs to create particle beams
that are essentially coherent matter waves, i.e., a ``matter
laser.'' Such a matter laser could form the ``working fluid'' of our
proposed gravimetric device. In particular, in analogy to laser
physics, the interference of two individual Bose-Einstein condensate
wavefunctions demonstrates multiparticle interference with matter
waves; see \cite{Anderson, Andrews, shin} and the references
therein. This body of work demonstrates that Bose condensed
molecules are ``laser like'': they are coherent and show long-range
correlations. Indeed, the first BEC was achieved with Rubidium atoms
in 1995 \cite{Anderson}, cooled to 170[nK]. Rubidium has an atomic
weight of approximately 85[au], so that, in order to achieve the
desired mass to velocity ratio for our device as described above, we
require a particle velocity of $1.4 \times 10^{-7}$[m/s]. Using the
de Broglie formula, this corresponds to a particle wavelength of
$\lambda = h/(mv) = 3.4$[cm]. Since this corresponds to the ground
state of a single-particle wavefunction, the condensate would have
to be created in a box with a length on the order of 1--10[cm].

In SI units, the critical temperature for condensate formation is
given by, $$T_c = 2.67 \times 10^{-45}\frac{n^{2/3}}{m}$$ where $n$
is the particle density and $m$ is the mass per particle. A
condensate temperature of 1[$\mu$ K] then requires a particle
density of $n = 4 \times 10^{20}$ [particles/$\rm{m}^3$] = $4\times
10^{14}$[particles/$\rm{cm}^3$]. For a condensate temperature of 1[n
K], which is on the order of the lowest achievable temperature to
date, a particle density on the order of
$10^9$[particles/$\rm{cm}^3$] is required. Given the required
dimensions of the container in which our BEC is to be created, this
means that it will be necessary to create a BEC with on the order of
a minimum of $10^9$ particles. Given that the largest BECs to date
have been achieved using on the order of $10^6$ particles [6], it is
clear that atom cooling and BEC technologies will have to be
developed some more before our proposed gravimetric device is
feasible.

Further, it has recently been experimentally demonstrated
\cite{shin} that it is possible to split gaseous Bose-Einstein
condensate into two coherent condensates by deforming an optical-
well where the condensate was trapped. Experiments analogous to what
is envisioned for the proposed device have been performed and the
two condensates were brought together at which point a matter-wave
interference pattern was observed. The coherent condensates were
separated for a duration of 5[ms] by 13[$\mu$m] and by 80[$\mu$m] in
\cite{shin,schumm}. The spatial scales in these experiments have
been several orders of magnitude smaller than the scales envisioned
for the proposed gravimeter. However, cryogenic technology exists,
and is developing fast, that is expected to make the necessary space
scales feasible. The temperatures that are needed to be maintained
along the path of the beam are of the order of 1[$\mu$K] while the
present-day record for achievable low temperatures is below 1[nK],
i.e., \emph{the record is more than two orders of magnitude lower
than the temperature that a gravitometer will require.} This is very
encouraging although the needed conditions will have to be
maintained for the substantial length that the matter wave would
need to traverse.

Finally, it should be emphasized that each gravimetric device will
actually consist of up to three interferometers, which will
separately measure the gravitational field in the $x, y,$ and $z$
directions, respectively. The one potential issue with using the $z$
direction is that the gravitational field in this direction is
already relatively strong, since it is the direction of the Earth's
gravitational field. A BEC or virus-based detector aligned along
this field may be too sensitive, and could essentially be
overwhelmed by the strength of the signal (i.e., the interference
pattern will be characterized by parameters that will make it
difficult to measure). Using an alternative working fluid, such as a
neutron beam, for this direction, will allow one to measure the
gravitational field in the z direction, however, it will not have
the required sensitivity to measure the extremely weak fluctuations
generated by a moving object. As a result, it may be preferable to
only measure the time-variation in the local gravitational fields in
the x and y directions. Then, instead of requiring two detectors per
moving object, it will be necessary to use three detectors.

\section{Signal Processing} \label{sec:signal}

Clearly, it is essential to determine whether or not our method for
detecting moving objects will work in the presence of various
disturbances that can introduce noise into the gravitational signal
at the detectors. There are two major sources of noise that will
need to be filtered out from the gravimetric signal. The first major
source of noise is due to moving objects that are not of interest,
such as civilian traffic and the movement of local wildlife. The
second is noise due to atmospheric disturbances. This includes
weather phenomena such as wind, precipitation, clouds, and even
simple convection patterns. There are several complementary
approaches to dealing with these sources of noise, all of which will
need to be considered in any feasibility study. First of all, a
moving object with a flight profile that would make it of interest
for detection will likely generate a gravimetric signal that has
distinct characteristics from the sources of noise described above.
Determining the defining features of this signal, along with the
defining features of the various sources of noise described above,
will allow us to design appropriate digital filters for extracting
the gravimetric signal generated by the moving object itself. In
particular, for certain types of moving objects, such as civilian
traffic, it may be possible to detect individual cars and airplanes
using optical or radar trackers. This could be done within a certain
radius of the detectors where their signal could be expected to
significantly distort the gravimetric signal generated by the object
to be detected. By accounting for the gravitational field generated
by these objects, it will be possible to filter out the contribution
that these objects make to the gravimetric signal at the detectors.
We will also explore general distributed models of civilian traffic
and wildlife, which is not directly detectable by other means, that
could affect the gravimetric signal, especially if the detectors are
placed next to major cities. Such civilian traffic should generate a
relatively constant signal with low frequency noise. In fact, one
should expect relative constancy of motion and slow speeds for many
typical non-military sources. Filtering out the effect of such
additional background sources from a typical military target should
be achievable quite effectively using standard bandpass filtering
techniques in the frequency domain. Next, it should be noted that,
for the case of general atmospheric disturbances, the overall effect
on the gravimetric signal is expected to be relatively weak. The
reason for this is that the gravitational field is determined
entirely by the mass distribution, and not by the velocity field
associated with the distribution. Therefore, even if there are
moving masses of air and precipitation, the overall change in the
mass distribution may be sufficiently small as to have a
comparatively small effect on the gravitational field.

The precise contribution of disturbances such as weather fronts and
wind gusts, albeit small, is expected to be more challenging. In
order to model such disturbances, we will need to work with fluid
dynamic models that describe atmospheric phenomena. For our
purposes, such models do not need to be overly sophisticated, as our
goal is not to predict specific weather patterns, but rather to
characterize the gravimetric signal generated by such phenomena, in
order to develop an appropriate filter to detect and remove them. An
appropriate class of such models has been proposed in the
meteorology literature based on the movements of weather fronts and
using the theory of mass transport \cite{Cullen}. These models are
very easy to implement on computer \cite{Eldad}, and could allow us
to make the necessary differentiation of the signature of a moving
(compressible) mass of air as opposed to that of a rigid object such
as an aircraft. We anticipate the spectral content of these two
signals to be significantly different, thereby allowing statistical
analysis and the relevant filtering techniques to be brought to bear
on the problem.

\section{Discussion of Numerical Methods}

We review some of the numerical methods that were employed in our
simulations of Section~\ref{sec:simulations}.

\newcommand{\vF}{\vec{F}}
\newcommand{\va}{\vec{a}}
\newcommand{\vx}{\vec{x}}

\subsection{Newton-Raphson Iteration} \label{sec:appendixB}

The goal behind Newton-Raphson Iteration, or simply Newton's Method,
is to solve a nonlinear system of equations $\vF(\va)=\vx$. The idea
behind the method is to choose an initial guess, denoted $\vx_0,$
that is reasonably close to the actual solution, which we denote by
$\vx_{{sol}}.$ If this is the case, then we may make a first-order
approximation and write, \begin{equation} \va = \vF(\vx_{sol}) =
 \vF(\vx_0) + D_{\vx}
\vF(\vx_0)(\vx_{{sol}} - \vx_0), \label{A1}\end{equation} and so we
may solve for $\vx_{{sol}}$ by solving the linear system $D_{\vx}
\vF(\vx_0)(\vx_{{sol}} - \vx_0) = \va - \vF(\vx_0).$

In reality, however, the value obtained for $\vx_{{sol}}$ using the
above procedure will not give the actual value for $\vx_{{sol}}$,
but some other value, denoted $\vx_1.$ The reason for this is that
the first-order Taylor expansion is not exact. Nevertheless, if
$\vx_0$ is close enough to $\vx_{{sol}}$ so that the first-order
Taylor expansion is sufficiently accurate, then at the very least
$\vx_1$ will be closer to the true value of $\vx_{{sol}}$ than
$\vx_0.$ This means that $\vx_1$ may be used as an improved initial
guess for Eq.~(\ref{A1}), which should then generate an improved
estimate for $\vx_{{sol}}$, denoted $\vx_2.$ Continuing in this way,
we generate a sequence $\{\vx_n\}$ of points, converging to
$\vx_{{sol}},$ that are related to each other by the recursion
relation, \begin{equation} D_{\vx} \vF(\vx_n)(\vx_{n+1} - \vx_n) =
\va - \vF(\vx_n). \label{A2}
\end{equation}

\subsection{Newton-Raphson Iteration with Solution Deformation}
\label{sec:appendixC}

Very often, the difficulty with obtaining convergence using Newton's
Method stems from an inability to pick a good initial guess. We
illustrate one approach for dealing with this problem: Suppose we
wish to solve the nonlinear system of equations $\vF(\vx) = \va,$
and we are given some initial guess $\vx_0.$ This initial guess will
not generate a sequence $\{\vx_n\}$ that converges to the desired
solution, nevertheless, this initial guess is essentially as good as
any other, since the system of equations is such that it is
difficult to choose a good initial guess. The idea is to therefore
start with the given initial guess $\vx_0$ and see if it is possible
to reach the desired solution to the system of equations. We begin
by defining $\va_0 = \vF(\vx_0),$ so that $\va_0$ is the value of
the function evaluated at the initial guess. We then define a
continuous curve $\vec{\alpha}(s)$ with the properties that
$\vec{\alpha}(0) = \va_0$ and $\vec{\alpha}(1) = \va.$ Thus, as $s$
ranges from 0 to 1, $\vec{\alpha}$ goes from $\va_0$ to $\va.$ We
now choose some positive integer $M,$ and for $m = 0, \ldots ,M,$ we
define $\va_m = \vec{\alpha}(m/M).$ Note that this generates a
sequence $\va_0, \va_1, \ldots, \va_M = \va.$ By continuity, the
distance between successive values of $\va_m$ decreases as $M$
increases.

So suppose we are able to obtain the solution, denoted $\vx_m$, to
the nonlinear system of equations $\vF(\vx) = {\va}_m.$ For large
$M,$ the difference between $\va_m$ and $\va_{m+1}$ should be fairly
small, so that $\vx_{m+1}$ should be fairly close to the solution to
the nonlinear system of equations $\vF(\vx) = {\va}_{m+1}.$
Therefore, if we use $\vx_m$ as the initial guess to the solution
for $\vF(\vx) = {\va}_{m+1}$, then Newton's Method is fairly likely
to converge in such a case.

In this way, starting from the initial guess $\vx_0,$ which is the
solution to $\vF(\vx) = {\va}_{0}$, we obtain from Newton's Method a
sequence of points $\vx_0, \vx_1, \ldots , \vx_M,$ where $\vF(\vx_m)
= \va_m.$ At each step, where Newton's Method is used to generate
$\vx_{m+1}$ from $\vx_m,$ convergence is likely, because $\va_m$ and
$\va_{m+1}$ are sufficiently close that $\vx_m$ and $\vx_{m+1}$ are
close enough as well for Newton's Method to converge using $\vx_m$
as the initial guess. Note that this approach does not attempt to
solve the original system $\vF(\vx) = \va.$ Rather, it starts with a
solution vector $\va_0$ for which $\vx_0$ is the solution to
$\vF(\vx) = \va_0,$ and then continuously deforms $\va_0$ into
$\va.$ As a result, we term this method Newton-Raphson Iteration
with Solution Deformation. Finally, although many deformations of
$\va_0$ into $\va$ are possible, here we employ a linear
deformation, given by $\vec{\alpha}(s) = (1 - s)\va_0 + s\va,$ where
$s \in [0,1],$ in our simulations.

\section{Conclusions and Future Research}

This paper describes a possible scheme for the construction of a
gravimetric radar. The goal is the detection of large, fast moving
objects, within a reasonable range, through the extraction of
signals from gravimetric data. This will have direct applications to
stealth technology. At present, feasibility of such a device hinges
on the development of sensors that are four orders of magnitude
better than existing technology. It is envisioned that such a device
will be based on gravity-induced quantum interferometry and the use
of Bose-Einstein condensates in the form of particle beams with
relatively massive, ultra-cold particles. The functionality and
reliability of the gravimetric radar will rely critically on a
substantial signal processing component to account and mediate the
effects of known disturbances produced by other large moving objects
or weather fronts.

\section*{Acknowledgement}

This research was partially funded by Boeing.

\end{document}